\documentclass[twocolumn,aps,showpacs,preprintnumbers,amsmath,amssymb,superscriptaddress,nobibnotes]{revtex4}

\usepackage{graphicx}
\usepackage{dcolumn}
\usepackage{bm}
\usepackage{graphics}
\usepackage{longtable}

\begin{document}

\title{Calibration of a single atom detector for atomic micro chips}

\author{A. Stibor}
\affiliation{Physikalisches Institut der
Universit\"{a}t T\"{u}bingen, Auf der Morgenstelle 14, D-72076
T\"{u}bingen, Germany}
\author{S. Kraft}
\altaffiliation[Current address: ] {van der Waals-Zeeman Instituut,
Universiteit van Amsterdam, Valckenierstraat 65, 1018 XE Amsterdam,
The Netherlands}
\affiliation{Physikalisches Institut der
Universit\"{a}t T\"{u}bingen, Auf der Morgenstelle 14, D-72076
T\"{u}bingen, Germany}
\author{T. Campey}
\affiliation{School of Physical Sciences, The University of
Queensland, Brisbane 4072, Australia}
\author{D. Komma}
\affiliation{Physikalisches Institut der Universit\"{a}t
T\"{u}bingen, Auf der Morgenstelle 14, D-72076 T\"{u}bingen,
Germany}
\author{A. G\"{u}nther}
\affiliation{Physikalisches Institut der Universit\"{a}t
T\"{u}bingen, Auf der Morgenstelle 14, D-72076 T\"{u}bingen,
Germany}
\author{J.~Fort\'{a}gh}
\affiliation{Physikalisches Institut der Universit\"{a}t
T\"{u}bingen, Auf der Morgenstelle 14, D-72076 T\"{u}bingen,
Germany}
\author{C. J. Vale}
\altaffiliation[Current address: ] {Centre for Atom Optics and
Ultrafast Spectroscopy, Swinburne University of Technology, Hawthorn
3122, Australia} \affiliation{School of Physical Sciences, The
University of Queensland, Brisbane 4072, Australia}
\author{H. Rubinsztein-Dunlop}
\affiliation{School of Physical Sciences, The University of
Queensland, Brisbane 4072, Australia}
\author{C. Zimmermann}
\affiliation{Physikalisches Institut der Universit\"{a}t
T\"{u}bingen, Auf der Morgenstelle 14, D-72076 T\"{u}bingen,
Germany}


\begin{abstract}
We experimentally investigate a scheme for detecting single atoms
magnetically trapped on an atom chip. The detector is based on the
photoionization of atoms and the subsequent detection of the
generated ions. We describe the characterization of the ion detector
with emphasis on its calibration via the correlation of ions with
simultaneously generated electrons. A detection efficiency of
47.8~$\pm$~2.6\% is measured, which is useful for single atom
detection, and close to the limit allowing atom counting with
sub-Poissonian uncertainty.
\end{abstract}

\maketitle

\section{Introduction}
Trapping and manipulating a small number of cold atoms in optical or
magnetic traps is one of the most intriguing fields of research with
ultra cold atoms. Degenerate quantum gases with small atom numbers
allow for the investigation of physics beyond the mean field
approach such as the Mott-Insulator transition \cite{Greiner2002a},
revival of coherence \cite{Greiner2002b}, number squeezing
\cite{chuu2005,Jo2006a,li2007}, Tonks-Girardau gases
\cite{Paredes2004a,Kinoshita2004a} and Luttinger liquids
\cite{Voit1995a,Wonneberger2006a}. Furthermore controlled quantum
engineering of few atom entanglement is a mandatory prerequisite for
quantum computation with cold atoms. The implementation of such
scenarios is one of the main motivations for trapping atoms in
miniaturized magnetic traps near the surface of a microchip
\cite{Zimmermann2007a}. In addition, such magnetic micro traps offer
unique opportunities for atom interferometry \cite{Jo2006a,
Schumm2005a, wang2005, Guenther2006a} and surface diagnostics
\cite{Wildermuth2005a, Guenther2005a}.

To date, micro traps are exclusively operated with ensembles such as
clouds of thermal atoms or quantum degenerate gases
\cite{Zimmermann2007a}. The ability to detect single atoms near the
surface of a chip would open the door to a new class of experiments.
Besides future applications in quantum computation \cite{birkl2007},
a continuous flux of atoms transmitted through an on-chip atom
interferometer could for instance be detected in real time while the
resonance of the interferometer is varied. A complete interferometer
spectrum could thus be recorded with atoms from a single
Bose-Einstein condensate. In combination with an integrated atom
laser \cite{fortagh2003a}, a sensitive matter wave based
spectrometer could be realized similar to its optical analog
involving tunable lasers and photo detectors.

Besides the well known experiments with single atoms in cavities
\cite{Pinkse2000a,Haroche2001a}, there has recently been significant
progress towards the detection of single cold atoms. In the special
situation of metastable helium, direct single atom detection with
micro channel plates is possible due to the large internal energy of
the atoms that is released at the surface of the detector
\cite{Schellekens2005a, Jeltes2006a}. Sensitive single atom
detection has also been demonstrated with optical cavities placed
either at some distance from the atom source \cite{Bourdel2006a} or,
very recently, integrated on an atom chip
\cite{Teper2006a,Treutlein2006a}. Alternatively, the atoms can be
ionized by optical excitation with the resulting ions being
subsequently detected with a suitable ion detector. Photoionization
of atoms in an atom chip trap has recently been demonstrated
\cite{Kraft2007a}. By introducing an efficient ion detector that is
compatible with the chip geometry, single atom detection is
possible.

In this paper we experimentally investigate an ion detection scheme
that is compatible with atom chip traps. The first step of single
atom detection i.e. the photoionization of atoms in the micro trap,
is described elsewhere \cite{Kraft2007a}. Here, we describe the ion
detection scheme and demonstrate a calibration procedure for a
single ion detector which involves ionizing a large number of atoms
and correlating the ions with the simultaneously generated
electrons. The experimentally observed ion detection efficiency of
about 50\% is limited mainly by the sensitivity of the channel
electron multiplier (CEM) used for ion detection. There are standard
methods to increase this sensitivity. The total sensitivity would
then exceed the critical single atom detection efficiency of 50\%
which marks the threshold above which total atom number
determination with sub-Poissonian resolution is in principle
possible \cite{campey2006a}.

\section{Ion detection scheme}

\begin{figure}
\includegraphics[width=3.4in]{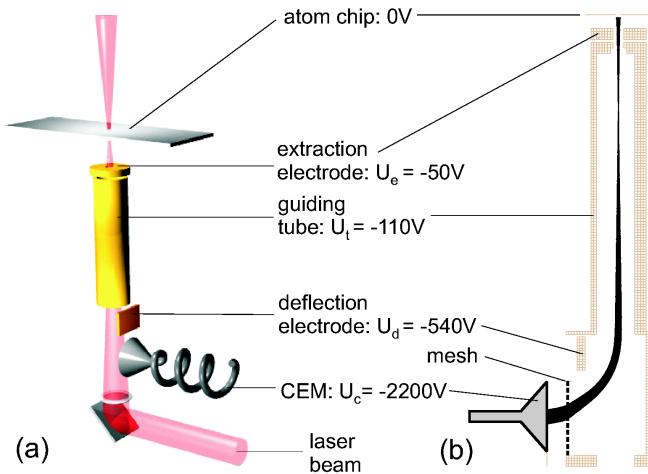} \caption{(Color online) (a) Detection scheme, showing the position
of the detector relative to the atom chip. The geometry allows ions
to be extracted from close to the chip surface. (b)~Numerical
simulation of the ion trajectories. With a suitable choice of
electrode voltages, ions within a well defined region can be guided
into the CEM with high efficiency, while ions outside this region
are not detected.} \label{fig:setup}
\end{figure}

The proposed ion detection scheme is shown in
Fig.~\ref{fig:setup}(a). Single atoms propagating in a magnetic
\mbox{waveguide} near the surface of an atom chip are photoionized
by stepwise excitation with two co-axial laser
beams~\cite{Kraft2007a}. The beams are oriented perpendicular to the
chip surface and pass through it via a 150~$\mu$m diameter hole.
This allows the detection of atoms at very small distances from the
surface of 1~$\mu$m and below. Ionization occurs within the
30~$\mu$m beam waist and the spatial selectivity of ionization could
potentially be reduced further to 100~nm \cite{Kraft2007a}. To allow
optical access to the trapped atoms and to create a well-defined
detection region, ion optics are used to guide the ions away from
the chip and into the CEM. To allow access for the ionization
lasers, the CEM is mounted at the end of the ion optics at
90$^\circ$ to the axis of the laser beams, and an additional
electrode deflects the ions into the CEM.

The first part of the ion optics consists of a disc (extraction
electrode) of 8~mm diameter which is placed parallel to the chip
surface at a distance of 1.6~mm. The chip is grounded and the
extraction electrode is biased at a negative voltage $U_e$, creating
an electric field in the region between them. This field accelerates
the ions from the ionization region towards a 1~mm diameter aperture
in the the extraction electrode, through which they enter a 36~mm
long tube set to a voltage $U_t$. The two electrodes together form a
lens which focuses the ions within the tube. At the end of the tube,
40~mm from the extraction electrode, a third electrode, set to
voltage $U_d$, deflects the ions, which leave the ion optics through
a fine mesh with 87\% transmission. Outside the ion optics they are
attracted by the horn of the CEM \cite{burle} set to a negative
voltage $U_c$. Any ions produced by the ionisation lasers within the
ion optics are screened from the CEM potential by the mesh. At the
surface of the CEM an ion impact causes the emission of secondary
electrons which starts an avalanche inside the multiplier tube. The
resulting charge pulse is observed with standard counting
electronics \cite{srs}.

Fig. \ref{fig:setup}(b) shows numerically simulated ion trajectories
for this setup. To a very good approximation, the ions move along
straight lines normal to the surface of the extraction electrode
such that the ion detection volume is simply given by the projection
of the extraction electrode aperture. This means that ions can only
be detected from the cylindrical region of diameter 1~mm running
between the aperture and the chip, effectively eliminating
background counts.

\section{Detector calibration}

In order to calibrate the ion detector we made use of a scheme in
which atoms are ionized and the resulting electrons and electron-ion
coincidences are detected and counted \cite{campey2006a,ncrp}. By
only counting ions for which the correlated electron is also
detected, it is possible to measure the ion detection efficiency
without any knowledge of the electron detection efficiency. If $N$
atoms are ionized, the number of electrons detected, corrected for
background electrons, is $N_e=\eta_e N$, where $\eta_e$ is the
electron detection efficiency. The number of electron-ion
coincidences detected, corrected for false coincidences, is $N_c =
\eta_i \eta_e N$, where $\eta_i$ is the ion detection efficiency.
Then,\begin{equation} \label{eq:eta1} \eta_i = \frac{N_c}{N_e}.
\end{equation}

\begin{figure}
\includegraphics[width=2.8in]{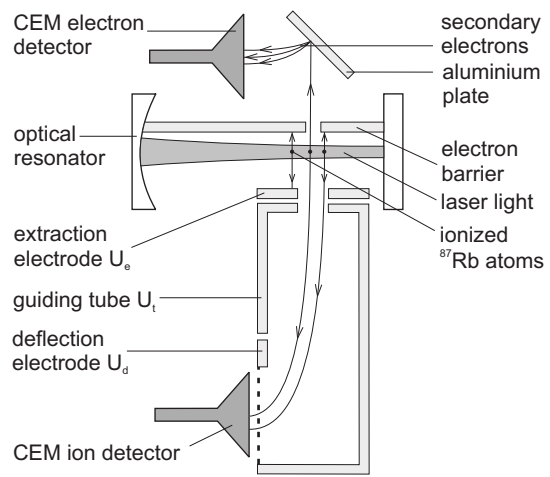} \caption{Setup for
calibration of the ion detector. Rubidium atoms from a background
vapor are photoionized within the mode of an optical resonator. The
generated electrons are observed with an additional CEM and the
ion-electron coincidence signal is also recorded.
}\label{fig:calsetup}
\end{figure}

The calibration setup is shown in Fig. \ref{fig:calsetup}. Ions are
produced by photoionizing $^{87}$Rb atoms within a standing wave
optical resonator. A commercial rubidium dispenser provides an even
rubidium pressure throughout the vacuum chamber containing the
apparatus. The background pressure, measured with an ion gauge, is
$2 \times 10^{-6}$ mbar. The optical resonator is located adjacent
to the ion optics with its optical axis parallel to the extraction
electrode. The resonator is comprised of a flat mirror and a curved
mirror (18 mm radius of curvature) separated by 12 mm, resulting in
a Gaussian fundamental mode with a beam radius of $w_0=$~46~$\mu$m.
Laser light is coupled into the resonator through the flat mirror
with an incoupling efficiency of 80\%. The light is taken from a cw
single mode diode laser system with a wavelength of 778.1066~nm,
resonant to the 5S$_{1/2} \rightarrow$ 5D$_{5/2}$ transition of
$^{87}$Rb. We observe a finesse of around 250 and a circulating
power of up to 1.5 W within the resonator. The cavity is
electronically locked to the side of the resonance peak by
controlling the mirror separation with a piezo element. The cavity,
including the two mirrors, is shielded from stray electric fields
(in particular from the piezo element).

Rubidium atoms that enter the resonator mode are excited from the
5S$_{1/2}$ to the 5D$_{5/2}$ state by two photon absorption which is
resonantly enhanced by the 5P$_{3/2}$ state. The transition is
Doppler free such that the atoms are excited independent of their
velocity. Under these conditions, the transition, which has a
natural linewidth below 500~kHz \cite{nez1993}, saturates for
circulating power greater than 5 mW. The population of the
5D$_{5/2}$ state is ionized by absorbing a third photon from the
resonator mode. The ionization rate is far from saturated due to the
short amount of time the atoms spend in the laser field. Together
with the strong saturation of the two photon transition, this
guarantees that the ionization rate grows linearly with the local
intensity. Consequently, the number of generated ions is constant
along the axis of the resonator and depends linearly on the
circulating power.

A grounded barrier, simulating the atom chip, is located on the
other side of the cavity mode from the ion optics. Electrons from
the ion detection region pass through the barrier via a 0.5 mm
diameter aperture. The fact that this aperture is smaller than the
ion extraction aperture, together with the straight field lines
between the barrier and the ion extraction electrode, ensure that no
electrons from outside the ion detection region are detected,
preserving the validity of Eq. \ref{eq:eta1}. The electrons that
pass through the aperture impact a grounded aluminium plate, from
which secondary electrons are detected with a second CEM with a horn
voltage of 300~V. The detection of secondary electrons was necessary
due to space limitations preventing mounting the CEM directly
adjacent to the aperture.

\section{Results}

\subsection{Optimization of electrode voltages}

\begin{figure}
\includegraphics[width=3.2in]{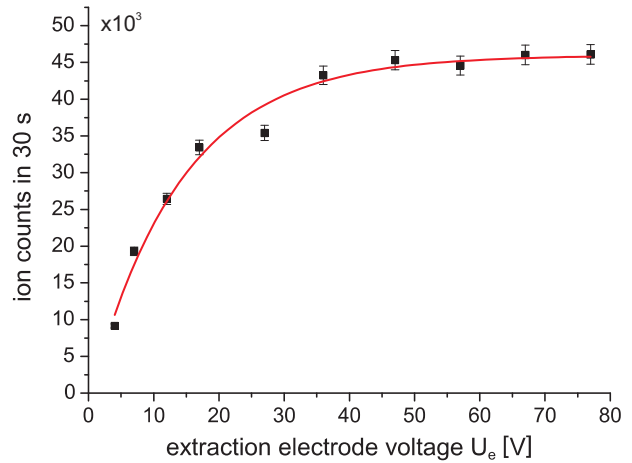} \caption{(Color online) Ion detection rate as a function of $U_e$, with
$U_t$ and $U_d$ set to their optimized values. Saturation of the
detection rate is observed for large values of $U_e$. The error bars
account for Poissonian fluctuations and the measured fluctations in
the ion detection rate of 2.9\%.}\label{fig:ionsat}
\end{figure}

The ion optics electrode voltages were optimized in order to
maximize the ion detection rate. For a given value of the extraction
electrode voltage $U_e$, the tube voltage $U_t$ and the deflection
electrode voltage $U_d$ were independently varied and the resulting
ion detection rate was measured. The maximized ion detection rate is
plotted in Fig. \ref{fig:ionsat} as a function of $U_e$, with $U_t$
and $U_d$ optimized. Since the ion optics contain only electrostatic
elements, in the absence of stray fields the ion trajectories, and
hence ion detection rate, depend only on the ratios of the electrode
voltages. In Fig. \ref{fig:ionsat} the ion detection rate shows an
initial increase, followed by a leveling off for values of $U_e$
greater than about 40 V. This indicates that below this voltage,
stray fields in the ionization region reduce the detection rate.

\begin{figure}
\includegraphics[width=3.2in]{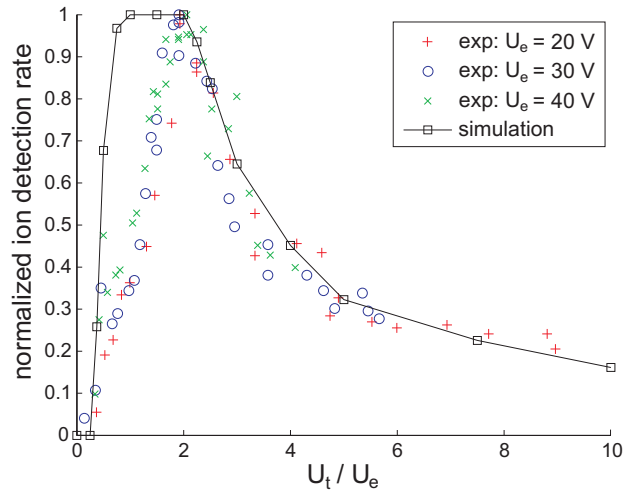} \caption{(Color online) Normalized ion detection rate as a function of $U_t/U_e$, with $U_d$
optimized. Experimental results for three different values of $U_e$
and results from the numerical simulation are
shown.}\label{fig:expvsim}
\end{figure}

Fig. \ref{fig:expvsim} shows the normalized ion detection rate as a
function of the ratio $U_t/U_e$,  which determines the focal length
of the ion lens. The deflection voltage $U_d$ was optimized for each
value of $U_t/U_e$. Experimental results are shown for three
different values of $U_e$ and show good agreement with each other,
consistent with the fact that the normalized ion detection rate
depends on the ratios between the electrode voltages, and not on
their absolute values. The curve predicted from the simulation is
also shown, and agrees broadly with the experimental results. The
difference for low values of $U_t/U_e$ is probably due to small
variations between the actual and the simulated geometry. In
particular, due to the fact that the atoms are ionized less than
1~mm from the electron barrier, the ion trajectories are very
sensitive to the position of the hole in the electron barrier
relative to the ion optics.

\begin{figure}
\includegraphics[width=3.2in]{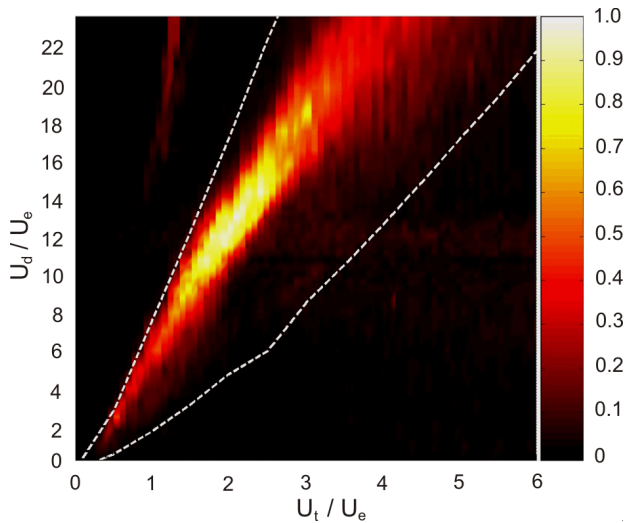} \caption{(Color online) Normalized ion detection rate as a function of
$U_t/U_e$ and $U_d/U_e$. The region between the dashed lines is the
region predicted by the numerical simulations for which ions impact
the CEM.}\label{fig:contour}
\end{figure}

The full dependence of the normalized detection efficiency on the
ratios between the electrode voltages is shown in Fig.
\ref{fig:contour}. This data is for $U_e=40$~V, although essentially
the same dependence was observed for other voltages. A non-zero
count rate is obtained for relatively large variations in $U_t/U_e$
and $U_d/U_e$ from their optimal values. Also shown on the plot is
the region for which ions impact the CEM, as predicted by numerical
simulations. The large region of non-zero count rate and the good
agreement between the simulations and experiment means that it is
straightforward to set the electrode voltages to values that produce
an ion signal. This is an important consideration in setting up an
on-chip experiment, in which ions are produced by ionizing atoms
from a BEC. This represents a pulsed source with relatively small
ion numbers, in contrast to the current situation in which ions are
produced by a high flux continuous source.

\subsection{Coincidence spectrum}

\begin{figure}
\includegraphics[width=3.2in]{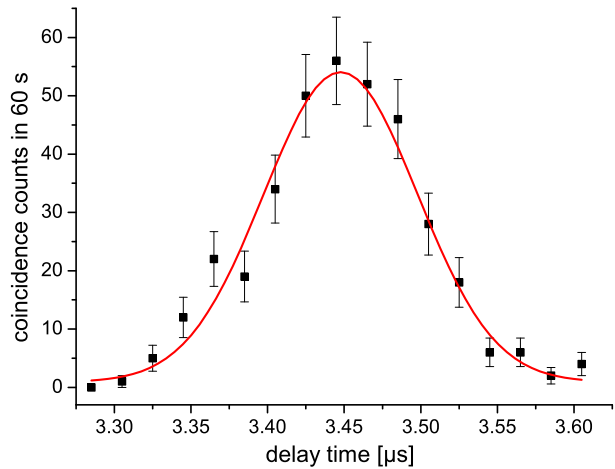} \caption{(Color online) Coincidence spectrum showing the
number of electron-ion coincidences counted in 60~s within a 20~ns
coincidence window as a function of the delay time of that window.
The error bars correspond to the square roots of the data, assuming
the detection of coincidences follows Poisson statistics. A Gaussian
fit is also shown (solid line) and has a mean value of 3.45~$\mu$s
and a standard deviation of 50~ns.} \label{fig:coinc}
\end{figure}

Due to the low mass to charge ratio of electrons relative to the
ions, following the ionization of an atom, the electron is detected
before the ion. Coincidences were therefore detected by using the
electron pulses to trigger a time window for the detection of ion
pulses. In order to define this window, it was necessary to measure
the coincidence spectrum for the delay times between ions and
electrons. This was done by counting the ions arriving within a
narrow time window (of width 20~ns) triggered by each electron pulse
and following a certain delay time. For a given delay time, the
number of ions detected in the 20~ns window was counted over 60~s.
By scanning the delay time of the window, the coincidence spectrum
was generated (Fig. \ref{fig:coinc}). A peak of true coincidence is
observed against a background of false coincidences. The mean delay
time given by a Gaussian fit was 3.45~$\mu$s and the standard
deviation was 50~ns. Within the calibration experiments,
coincidences were counted using an 800 ns window centered on the
peak of the coincidence spectrum.

\subsection{Ion detection efficiency}

To measure the efficiency of the ion detector, calibration
experiments were carried out with optimized voltages of the ion
optics electrodes. In each experiment the numbers of electrons and
coincidences were counted in periods of 100~s. The average number of
background electrons was measured in separate 100~s periods with the
laser detuned 44~GHz off-resonance. The average number of false
coincidences was measured by counting the number of coincidences in
100 s periods, with the coincidence window displaced 4~$\mu$s from
the peak of the coincidence spectrum. For our system, the dead time
following the detection of a pulse is of order 50 ns, which for the
count rates used has a negligible impact on the measured ion
detection efficiency.

\begin{figure}
\includegraphics[width=3.2in]{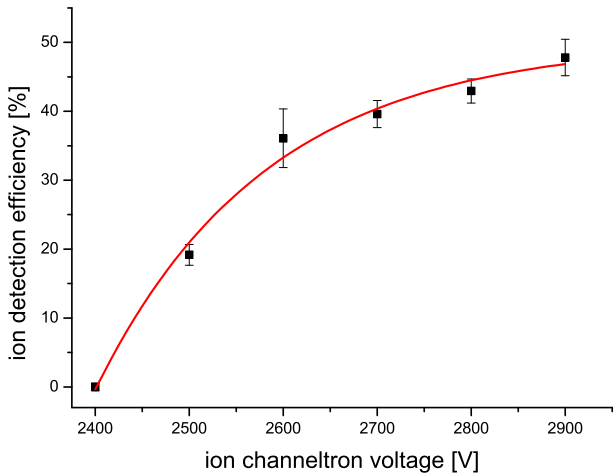} \caption{(Color online) The ion detection efficiency as a function of the voltage
 across the ion CEM. Saturation at about 50\% is observed for large values of the CEM voltage.
} \label{fig:CHsaturation}
\end{figure}

The ion detection efficiency was measured as a function of the
voltage across the CEM.  For each voltage, several calibration
experiments were carried out and the ion detection efficiency for
each experiment was calculated from Eq. \ref{eq:eta1}. The curve of
the ion detection efficiency was observed to saturate near 50\%
(Fig. \ref{fig:CHsaturation}). The highest voltage at which we
measured the ion detection efficiency was limited by the CEM
specifications to 2900~V. At this voltage, the mean value of the ion
detection efficiency was $\eta_i=47.8\%$ and the standard deviation
was $2.6\%$.

\section{Discussion}

The efficiency of our detector lies at the threshold of the regime
in which sub-Poissonian counting is possible. The ion detection
efficiency we measured is valid for ions originating from the region
from which electrons were also detected. This is due to the fact
that only ions correlated with detected electrons were counted in
the calibration experiments. According to our ion optics
simulations, all the ions from this region are guided to the CEM.
Therefore, apart from transmission through the mesh in front of the
CEM, the collection efficiency of the ion optics is not a limiting
factor in the overall ion detection efficiency. The overall
efficiency is then limited by the efficiency of the CEM. This is
mainly due to secondary electron emission within the CEM. The number
of secondary electrons emitted by an ion impact is approximately
described by a Poisson distribution \cite{dietz1975}. With the ion
energies and CEM material used here, there is a significant
probability that no secondary electrons result from an ion impact,
or that the electron cascade dies out in the channel. The efficiency
could be increased by using a conversion dynode made of a material
with a high secondary electron emission coefficient and set to a
high voltage. Efficiencies close to 100\% may be achieved in
principle \cite{dietz1975}. However, the required voltages of order
10~kV present a technical challenge in regards to their
implementation in an atom chip setup.

In addition to a high detection efficiency, sub-Poissonian atom
counting requires low uncertainty in the detection efficiency
\cite{campey2006a}. This uncertainty can result from a lack of
knowledge of the detection efficiency, due to the binomial
statistics involved in the calibration, and from technical
fluctuations. The uncertainty of the ion detection efficiency due to
the binomial counting statistics and to the uncertainties in the
background electron and false coincidence rates was calculated
according to \cite{campey2006a} and found to be 1.7\%. The measured
value of 2.6\% has an uncertainty of 0.4\%, so while it is possible
that the fluctuations in the measured values of the detection
efficiency result only from the counting statistics, it is likely
that technical fluctuations are also present. These fluctuations
have a number of possible sources. Variations in the pulse height
distribution of the CEM pulses can cause changes in the percentage
of the CEM pulses that are above the discriminator level of the
pulse counting electronics. Variations in noise on the CEM signals
can also change the rate of background counts. Also, the detection
efficiency of a CEM depends on the position at which ions impact
\cite{gilmore1999}. Therefore, noise and drifts in the ion optics
electrode voltages which cause small changes in the ion trajectories
can introduce further fluctuations.

\section{Conclusions}

We have presented a detection scheme for single ultracold $^{87}$Rb
atoms in a magnetic microtrap on an atom chip. In this scheme, the
atoms are photoionized on the chip and the resulting ions are guided
by ion optics from near the chip surface to a CEM. Using a low
vacuum test chamber we were able to optimize and characterize the
detector. In this setup the atoms were ionized in a cavity and the
ions were collected by the ion optics. By using an additional CEM to
count electrons, and by counting electron-ion coincidences with the
ion CEM, we were able to measure an ion detection efficiency of 47.8
$\pm$2.6\%. This high detection efficiency make this detector
suitable for use in single atom on-chip interferometry and allows
the measurement of atom number correlations.

An additional feature of this detector is that when used with pulsed
ionization it represents a time of flight mass spectrometer
\cite{Wiley1955}. The detector has sufficient time resolution to
distinguish between ultracold atoms and molecules
\cite{sdkraft2006}, and thus would prove useful for experiments
involving mixtures of different species on atom chips.

From our ion optics simulations, the ion detection efficiency is
mainly limited by the efficiency of the CEM. Further improvements
can therefore be made by using a conversion dynode for signal
amplification. Despite the additional technical challenges this
presents, with such an improvement atom counting with sub-Poissonian
uncertainty and atom number squeezing experiments are feasible
\cite{campey2006a}. This, together with the detector's high single
atom detection efficiency and high spatial resolution, makes it a
sophisticated tool for probing degenerate quantum gases with small
particle numbers on atom chips, and allows the investigation of
physics beyond the mean field approximation.

\section{Acknowledgements}

This work was supported by the Landesstiftung Baden-W\"{u}rttemberg
through the project ``Atomoptik'', by the European Commission
through the Marie-Curie Research Training Network ``Atom Chips''
(MRTN-CT-2003-505032) and by the Australian Research Council. We
acknowledge helpful discussions with Herwig Ott and Sebastian Slama.

\end{document}